\documentclass[hackGuillaume]{svjour3} % hackGuillaume natbib
\usepackage[utf8]{inputenc}
\usepackage[T1]{fontenc}
\usepackage[english]{babel}
\usepackage{times,mathptmx}
\usepackage{latexsym} % $\rhd$
\usepackage{amsmath,amssymb}
\usepackage{nicefrac}
\usepackage{relsize}

\usepackage{booktabs}

\usepackage{graphicx}
\usepackage[usenames,svgnames,table]{xcolor}
\usepackage[colorlinks,allcolors=MediumBlue,breaklinks,hyperfootnotes=true,
	pdftitle={Sneaked references: Cooked reference metadata inflate citation counts},
	pdfauthor={Lonni Besançon, Guillaume Cabanac, Cyril Labbé, Alexander Magazinov},
	pdfsubject={Preprint},
	pdfkeywords={sneaked references, undue citations, citation manipulation, metadata registration, bibliometrics, research evaluation}
 ]{hyperref}

\usepackage{fontawesome5}    % \faCheck, \faSkullCrossbones
\usepackage{halloweenmath}   % \mathghost

\newcommand{\OK}{{\smaller\faCheck}}
\newcommand{\SN}{{\smaller$\boldsymbol{\mathghost}$}}
\newcommand{\LO}{{\smaller\faSkullCrossbones}}

\usepackage{tcolorbox} %tbox
\usepackage{hhline}
\usepackage{soul}
\usepackage{tikz}
\usetikzlibrary{backgrounds}
\usetikzlibrary{shapes,shadows,calc,arrows}
\usetikzlibrary{chains,matrix,positioning,scopes,snakes}
\usetikzlibrary{decorations.pathmorphing,shadows.blur,shadings}
\usetikzlibrary{decorations.pathreplacing,calc,fit}
\usetikzlibrary{arrows}
\usepackage[natbibapa]{apacite}    % instead of passing natbib to svjour3
\newcommand{\doi}[1]{\href{https://doi.org/#1}{#1}}
\usepackage{underscore}
\begin{document}
\journalname{}
\title{Sneaked references: Cooked reference metadata inflate citation counts}

\author{Lonni Besançon \and Guillaume Cabanac \and Cyril Labbé \and Alexander Magazinov}
\authorrunning{L. Besançon et al.}

\institute{
L. Besançon\at
Media and Information Technology, Linköping University, Norrköping, Sweden\\
\email{lonni.besancon@gmail.com}\\
ORCID: \href{https://orcid.org/0000-0002-7207-1276}{0000-0002-7207-1276}
\and
G. Cabanac\at
Université Toulouse~3 -- Paul Sabatier, IRIT UMR 5505 CNRS, 31062 Toulouse, France\\
Institut universitaire de France (IUF), France\\
\email{guillaume.cabanac@univ-tlse3.fr}\\
ORCID: \href{https://orcid.org/0000-0003-3060-6241}{0000-0003-3060-6241}
\and
C. Labbé\at
Univ. Grenoble Alpes, CNRS, Grenoble INP, LIG, 38000 Grenoble, France\\
\email{cyril.labbe@univ-grenoble-alpes.fr}\\
ORCID: \href{https://orcid.org/0000-0003-4855-7038}{0000-0003-4855-7038}
\and
A. Magazinov\at
Yandex.Kazakhstan, 43 Dostyq av., Almaty 050010, Kazakhstan\\
\email{magazinov-al@yandex.ru}\\
ORCID: \href{https://orcid.org/0000-0002-9406-013X}{0000-0002-9406-013X}
}

% Started writing on 28-MAR-2023
\date{Version: \today}
\maketitle 

\urlstyle{same}

\begin{abstract}
    We report evidence of an undocumented method to manipulate citation counts involving `sneaked' references.
    Sneaked references are registered as metadata for scientific articles in which they do not appear.
    This manipulation exploits trusted relationships between various actors: publishers, the Crossref metadata registration agency, digital libraries, and bibliometric platforms.
    By collecting metadata from various sources, we show that extra undue references are actually sneaked in at Digital Object Identifier (DOI) registration time, resulting in artificially inflated citation counts.    
    As a case study, focusing on three journals from a given publisher, we identified at least 9\% sneaked references ($\nicefrac{5,978}{65,836}$) mainly benefiting two authors. 
    Despite not existing in the articles, these sneaked references exist in metadata registries and inappropriately propagate to bibliometric dashboards.
    Furthermore, we discovered `lost' references: the studied bibliometric platform failed to index at least 56\% ($\nicefrac{36,939}{65,836}$) of the references listed in the HTML version of the publications.
    The extent of the sneaked and lost references in the global literature remains unknown and requires further investigations.
    Bibliometric platforms producing citation counts should identify, quantify, and correct these flaws to provide accurate data to their patrons and prevent further citation gaming.
\end{abstract}


%=========================================================================================================================
\section{Introduction}\label{sec:intro}

It is now well recognised that the \emph{Publish or Perish} atmosphere fuels questionable research practices \citep{CROUS:2019:DSA}. 
The introduction and widespread adoption of computed indicators ($h$-index, impact factor\ldots) has been leading academics to a situation where publishing is not enough and being cited is crucial.
In this world of \emph{Be Cited or Perish}, motivations for citation manipulations are on the rise \citep{lawrence2007mismeasurement}. 
Possibilities of such manipulations have been documented by whistleblowers and researchers alike \citep{baccini2019citation,haley2017inauspicious}.

\citet{BeelAndGipp2010} experimented hiding citations to human eyes by using `white on white' text. 
\citet{Labbe2010} achieved $h$-index manipulation through injection of meaningless texts containing a fixed set of references.
\citet{LopezCozarEtAl2014} reproduced the previous experiment, demonstrating how the $h$-index and impact factors of real researchers and journals can be manipulated.
It is worth noting that some editorial practices may not be too far away from this type of manipulation: a seemingly legitimate editorial could cite all articles from a journal, thereby increasing its Impact Factor \citep[e.g., ][]{FoleyAndValkonen2012,HeathersAndGrimes2022}. 
Another method is the so-called `citation cartel' method \citep{Franck1999}. 
As part of the cartel, you cite specific authors that will cite you in return. 
This kind of manipulation also arises at the journal level~\citep{Davis2016,Kojaku2021}. 
Another example is called `citation plantation'\footnote{\url{https://pubpeer.com/search?q=\%22citation+plantation\%22}} and refers to undue over-citation of certain authors, even on unrelated topics. 
Last but not least, one of the most famous and common methods, is the addition of references through the peer-review process. 
At review time, authors may be asked by reviewers and editors to add undue references to their submission. 
Whistleblowers and academic sleuths often try to detect citation manipulations through skews in citation (or self-citation) data~\citep{WrenAndGeorgescu2022,van2020signs,SzomszorEtAl2020}. 

As the motivation for and practice of citation manipulation gain traction, the consequences of such a practice are starting to become visible in academia.
From time to time, highly cited researchers are banned from editorial boards \citep{van2020highly} because of their unethical practice to trade citations for manuscript acceptance. 
In 2021, Clarivate excluded 300 researchers from its \emph{Highly Cited Researchers} list, and about 550 in 2022 \citep{Oransky2022}. 
This decision was taken based on evidence of citation manipulation. 
Another example: some malevolent individuals forge hijacked journals by imitating current or defunct journals \citep{AbalkinaEtAl2022a}. 
They publish non-reviewed papers that cite papers generating potential undue citations. 
Some manage to get these indexed by Elsevier's Scopus, a bibliometric platform that computes author-level indicators for research assessment \citep{BaasEtAl2020}.
    
It is worth pointing out that citation manipulation by various actors occurs at many places and at different times during the life cycle of a scientific publication. 
Up until now, the documented manipulations always implied modifications of the version of record \citep{Hinchliffe2022} (i.e., the \emph{real} article available in PDF/HTML in its final version) by adding references to it.
In this paper, we document a new flaw that is currently exploited: \emph{sneaking undue references} during the DOI registration by supplying extra and irrelevant metadata. 
The scientific publication itself, namely the version of record, remains unaltered and undue citations are actually \emph{unreachable} by readers.
We provide evidence that this manipulation is in use as we discovered in at least three journals of an open access publisher.
This exploit will remain available as long as the metadata pushed by publishers are not carefully verified.

%=========================================================================================================================
\section{The exploit: Increased citation counts with sneaked references}\label{sec:exploit}

From a paper's bibliography to bibliometric dashboards, the path is long for references to be counted.
Different actors using various deception techniques can sneak undue references in along this path. 

%-------------------------------------------------------------------------------------------------
\subsection{Context: the DOI and metadata registration process}
\label{sec:path}

\begin{figure}[!ht]
    \input{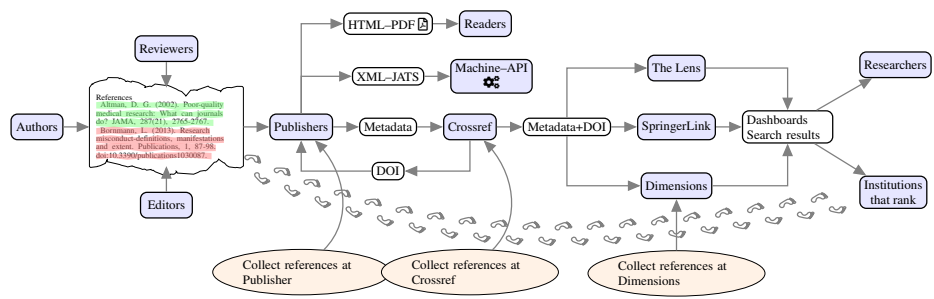}

%\usetikzlibrary{decorations.fractals}
\usetikzlibrary{decorations.footprints}

\resizebox{\columnwidth}{!}{

\begin{tikzpicture}
%above right = -4ex and 15em of KDB
% define some styles
\tikzstyle{actor} = [rectangle, draw, black,
                      minimum width=4em, minimum height=2em,rounded corners,align=center, fill=blue!10, thick]

\pgfdeclaredecoration{irregular fractal line}{init}
{
  \state{init}[width=\pgfdecoratedinputsegmentremainingdistance]
  {
    \pgfpathlineto{\pgfpoint{random*\pgfdecoratedinputsegmentremainingdistance}{(random*\pgfdecorationsegmentamplitude-0.02)*\pgfdecoratedinputsegmentremainingdistance}}
    \pgfpathlineto{\pgfpoint{\pgfdecoratedinputsegmentremainingdistance}{0pt}}
  }
}

\tikzset{
   paper/.style={draw=black, fill=white%blur shadow, %shade=linear interpolation,
                 %lower left=white, upper left=white, upper right=white, lower right=black!20},
                 %lower left=white, upper left=white, upper right=white, lower right=white
                 },
   irregular border/.style={decoration={irregular fractal line, amplitude=0.2},
           decorate,
     },
   ragged border/.style={ decoration={random steps, segment length=7mm, amplitude=2mm},
           decorate,
   }
}

%%%%%%%%%%%%%%%%%%%%%%%%%%%%%%%%%%%%%%% Part 1 %%%%%%%%%%%%%%%%%%%%%%%%%%%%%%%%%%%%%%%%%%%%%%%%%%
%            [reviewers]               
%                |                     
%                V                     
%[Authors] --(article)*--> [publisher] ---- (HTML)-----> readers  
%                \^                     ---- (PDF) -----> readers
%                |                     ---- (XML-JATS) ----> machine / API
%            [editors]

\node [actor] (rev) [] {Reviewers};
%%%%%%%
% Article
  \node[inner sep=0.5em,text width=0.25\textwidth] (art) [below =2em of rev] {
  \parbox{\linewidth}{\baselineskip=1pt
  {\tiny 
References\\
\highlight[green!50]{ Altman, D. G. (2002). Poor-quality medical research: What can journals do?
JAMA, 287(21), 2765-2767.
}
\\ 
\highlight[red!50]{
Bornmann, L. (2013). Research misconduct-definitions, manifestations and extent. Publications, 1, 87-98. doi:10.3390/publications1030087.
}
}}
};  % Draw the text of the node
\begin{pgfonlayer}{background}  % Draw the shape behind
\fill[paper] % recursively decorate the bottom border
      { decorate[irregular border]{decorate{decorate{decorate{decorate[ragged border]{
        (art.north west) -- (art.north east)
      }}}}}} -- (art.south east)
      {decorate[irregular border]{decorate{decorate{decorate{decorate[ragged border]{
      -- (art.south west)
      }}}}}}
      -- (art.north west);% (A.north west) -- cycle;  
\end{pgfonlayer}
% End Article
%%%%%%%

\node [actor] (aut) [left =2em of art]  {Authors};
\node [actor] (edit) [below =2em of art]  {Editors};

\node [actor] (pub) [right =2em of art]  {Publishers};

\node [draw,rounded corners,thick] (meta) [right =2em of pub]  {Metadata};
\node [draw,rounded corners,thick] (doi) [below =2em of meta]  {DOI};
\node [actor] (cros) [right =2em of meta]  {Crossref};

\node [draw,rounded corners,thick] (jats) [above =2.5em of meta]  {XML--JATS};
\node [actor] (mach) [right =2em of jats]  {Machine--API \\ \faCogs};

\node [draw,rounded corners,thick] (HTML) [above = 2.5em of jats]  {HTML--PDF \faFilePdf[regular]};
\node [actor] (read) [right =2em of HTML]  {Readers};

\draw[gray, thick, ->,>=triangle 45] (aut) -- (art);
\draw[gray, thick, ->,>=triangle 45] (art) -- (pub);
\draw[gray, thick, ->,>=triangle 45] (rev) -- (art);
%\draw[gray, thick, ->,>=triangle 45] (art) -- (pub);
\draw[gray, thick, ->,>=triangle 45] (edit) -- (art); % -- (pub);

\draw[gray, thick, ->,>=triangle 45] (pub) |- (HTML);
\draw[gray, thick, ->,>=triangle 45] (pub) |- (jats);

\draw[gray, thick, ->,>=triangle 45] (jats) -- (mach);
\draw[gray, thick, ->,>=triangle 45] (HTML) -- (read);

\draw[gray, thick, ->,>=triangle 45] (meta) -- (cros);
\draw[gray, thick, ->,>=triangle 45] (pub) -- (meta);
\draw[gray, thick, ->,>=triangle 45] (cros) |- (doi);
\draw[gray, thick, ->,>=triangle 45] (doi) -| (pub);

%\draw[gray, thick] (-1,2) -- (2,-4);
%\draw[gray, thick] (-1,-1) -- (2,2);
%\filldraw[black] (0,0) circle (2pt) node[anchor=west]{Intersection point};
%%%%%%%%%%%%%% Part 2
%
%            [reviewers]               ---- (HTML)-----> readers  
%                |                     ---- (PDF) -----> readers                                           
%                V                     ---- (XML-JATS) ----> machine / API                                 --(search results)--> researchers
%[Authors] --(article)--> [publisher] --(registration form)--> [Crossref] --(metadata+DOI)--> [Dimensions] --(dashboards)--> institutions
%                ∧                     <--------(DOI)----------                           --> [Lens]
%                |                                                                        --> [SpringerLink]
%            [editors]

\node [draw,rounded corners,thick] (metdo) [right = 2em of cros]  {Metadata+DOI};
\node [actor] (dim) [right =2em of metdo]  {SpringerLink};
\node [actor] (lens) [above =2.5em of dim]  {The Lens};
\node [actor] (spri) [below =2.5em of dim]  {Dimensions};

\node [draw,rounded corners,thick,text width=14ex] (hub) [right =2em of dim]  {Dashboards Search results};

\node [actor] (res) [above right =2.5em and 2em of hub]  {Researchers};
\node [actor] (inst) [below right =2.5em and 2em of hub]  {Institutions \\ that rank};

\draw[gray, thick, ->,>=triangle 45] (cros) -- (metdo);
\draw[gray, thick, ->,>=triangle 45] (metdo) -- (dim);
\draw[gray, thick, ->,>=triangle 45] (metdo) |- (lens);
\draw[gray, thick, ->,>=triangle 45] (metdo) |- (spri);
\draw[gray, thick, ->,>=triangle 45] (lens) -| (hub);
\draw[gray, thick, ->,>=triangle 45] (dim) -- (hub);
\draw[gray, thick, ->,>=triangle 45] (spri) -| (hub);

\draw[gray, thick, ->,>=triangle 45] (hub) -- (res);
\draw[gray, thick, ->,>=triangle 45] (hub) -- (inst);

%%%%%%%%%%%%%% Part 3
%
%            [reviewers]               ---- (HTML)-----> readers  
%                |                     ---- (PDF) -----> readers                                           
%                V                     ---- (XML-JATS) ----> machine / API                                 --(search results)--> researchers
%[Authors] --(article)--> [publisher] --(registration form)--> [Crossref] --(metadata+DOI)--> [Dimensions] --(dashboards)--> institutions
%                ^              ^       <--------(DOI)----------   ^                        --> [Lens]
%                |              |                                  |                        --> [SpringerLink]
%            [editors]          |                                  |                             ^
%                               |                                  |                             |
%                        Collect metadata                    Collect metadata                Collect metadata
%                from the Publisher’s web site               from Crossref                   from Dimensions

%text width=2cm, 
%{footprints,foot length=10pt,stride length=20pt}
\node [text width=20ex, fill=orange!10,draw,ellipse] (col1) [below =8.5em of pub]
        {Collect references at Publisher };
%decorate, decoration=saw,
\node [text width=20ex, fill=orange!10,draw,ellipse] (col2) [below =8.5em of cros]
        {Collect references at Crossref };

\node [text width=20ex, fill=orange!10,draw,ellipse] (col3) [below =4em of spri]
        {Collect references at Dimensions };

%gray, thick, --, decoration={footprints,foot length=10pt,stride length=20pt}]
\draw[gray, thick,decorate, decoration={footprints,foot length=10pt,stride length=20pt}] (art) to [out=-25,in=190] (inst);
%\fill decorate{ (art) to [bend left] (inst)}

\draw[gray, thick, ->,>=triangle 45] (col1) to [out=45,in=-45] (pub); 
\draw[gray, thick, ->,>=triangle 45] (col2) to [out=45,in=-45] (cros); 
\draw[gray, thick, ->,>=triangle 45] (col3) -- (spri); 

\end{tikzpicture}

} %end resize box
    \caption{References' long path from authors to bibliometric dashboards: after Editorial and Peer-Review assessment, metadata are registered to a DOI provider (here Crossref). Metadata are then retrieved by bibliometric platforms (The Lens, SpringerLink, Dimensions) that provide various services, such as a search engine and bibliometric dashboards for institutions.}\label{fig:walk}
\end{figure}

As \autoref{fig:walk} shows, after acceptance and before publication of papers, publishers register DOIs to Registration Agencies.
The main one is Crossref that mints DOIs for a fee and hosts the publishers' metadata that become publicly available~\citep{HendricksEtAl2020}. 
Most publishers push the reference lists of their papers as part of the registered metadata~\citep{SinghChawla2022}.\footnote{\url{https://www.crossref.org/documentation/schema-library/markup-guide-metadata-segments/references/}} 
Crossref is then used as a source by multiple platforms such as SpringerLink,\footnote{\url{https://citations.springernature.com/about}} The Lens~\citep{Penfold2020}, or Dimensions~\citep{HerzogEtAl2020}.\footnote{To the best of our knowledge, Google Scholar relies on various sources and crawling methods \citep{vanNoorden2014}.} 
Bibliometric platforms source from the metadata registered at Crossref \emph{inter alia} to report indicators at the individual/institutional/journal levels, such as citation counts, impact factors, and $h$-indices.

%-------------------------------------------------------------------------------------------------
\subsection{The manipulation \ldots{} explained}\label{sec:places}

Crossref hosts as-such the metadata sent by their members, namely the publishers:

\begin{quote}
    \begin{tcolorbox}[toprule=0pt,leftrule=1pt,rightrule=0pt,bottomrule=0pt]
        \smaller
        “Our metadata is provided to us by our members, and we don’t curate or clean up the metadata in any way. We do insert metadata into outputs such as DOI matches for citations, recursive relationships, and clearly flag those pieces as being inserted by Crossref in our metadata outputs.\\
        This means, good or bad, metadata accuracy depends on the quality of metadata provided by our members.”\\
        
        --- ‘Metadata principles and practices’ from Crossref\\\phantom{--- }\url{https://www.crossref.org/documentation/principles-practices/}
    \end{tcolorbox}
\end{quote}

When registering a new publication and its references at Crossref, a publisher may sneak extra undue references in the metadata sent in addition to the ones originally present.
Then, digital libraries (e.g., SpringerLink) and bibliometric platforms (e.g., Dimensions) harvest these metadata, undue citations included.
These sneaked references are processed and counted even if they are not present in the original publication.

This new way to manipulate citation counts relies on metadata manipulations that leave the original text untouched.
This exploit is made possible because Crossref trusts publishers to extract, report, and send them metadata about the publications, including the references.
As a matter of fact, Crossref not controlling the accuracy of the metadata provided by publishers creates a `security breach' within the information flow.
The next section shows that this manipulation is actually in use.

%=========================================================================================================================
\section{Case study: Evidence of sneaked references in three journals of a given publisher}\label{sec:case}

To provide evidence of citation counts manipulation, one needs to collect samples of metadata at three different places along the reference registration path depicted in \autoref{fig:walk}.
Sneaked references are revealed when comparing the reference lists of publications as provided 
1)~by the publisher on its website, 
2)~on the metadata registry at Crossref and 
3)~by a bibliometric platform: Dimensions.    

\begin{figure}[!ht]\centering
    \fbox{\includegraphics[width=\textwidth]{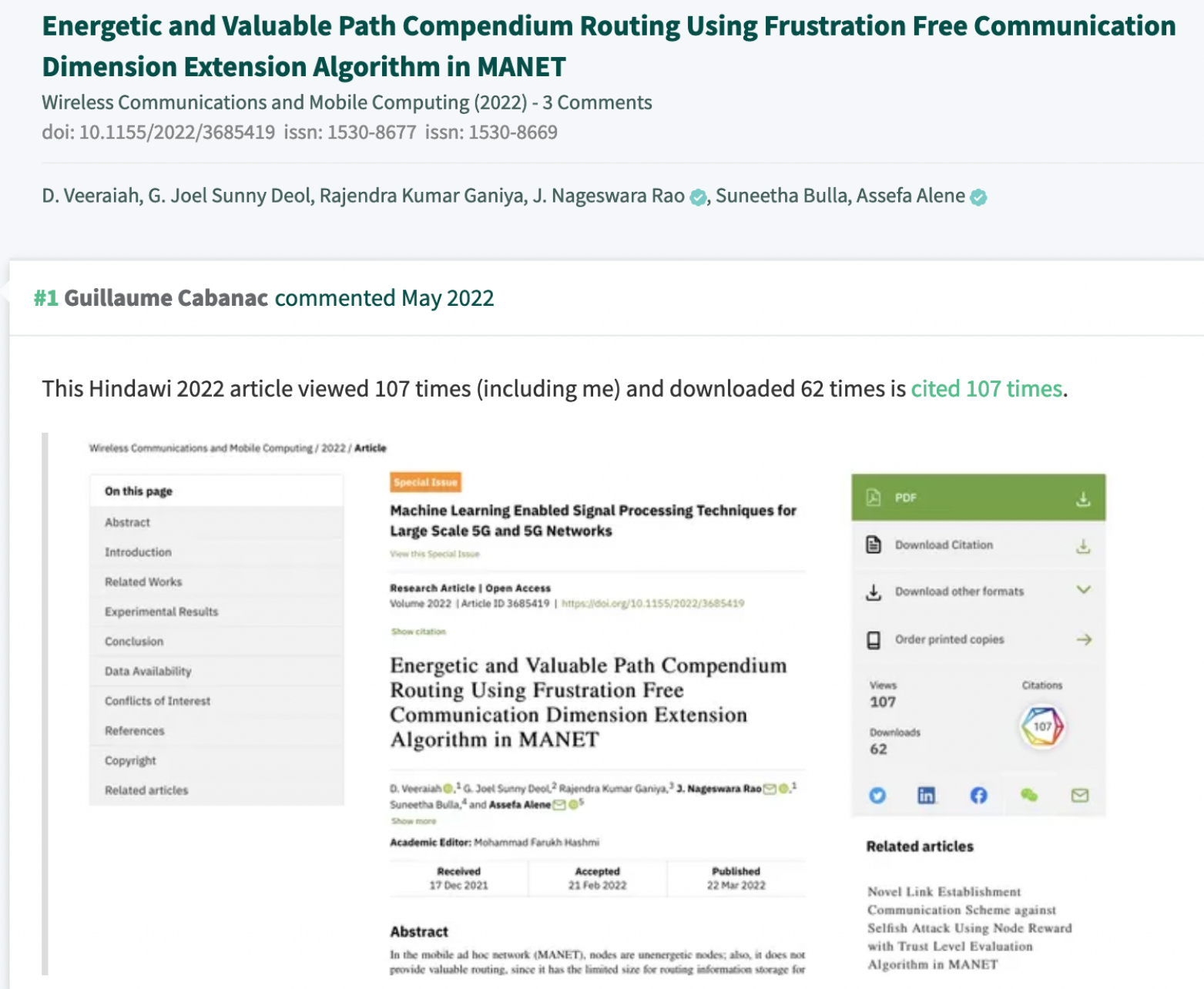}}
    \caption{PubPeer post \url{https://pubpeer.com/publications/A172115FC8D0A5F44B31A18B08BB26} reporting a Hindawi journal article with more citations than downloads. Most citations appear not to match any of the references in the allegedly citing publications. After careful examination, it appeared that these were sneaked references: existing in the metadata only and not in the PDFs of the allegedly `citing' publications.}
    \label{fig:PubPeerPost}
\end{figure}

As proof of the `sneaked references' manipulation happening, let us analyse three journals published by \emph{Technoscience Academy},\footnote{\url{https://technoscienceacademy.com}} an Indian open access publisher and Crossref member.
These three journals were selected after we identified incoherent metadata that we flagged in May 2022 on PubPeer (\autoref{fig:PubPeerPost}).
This case involves a Hindawi journal article published on 22 March 2022.
The Hindawi website showed a large number of citations ($n=107$) for a publication that had been online for less than two months.
On the screenshot in \autoref{fig:PubPeerPost}, the number 107 stems from Altmetric, a service offered by Dimensions that sources data from publishers and Crossref.\footnote{
    Crossref reported 107 citations for this paper, see the attribute \texttt{is-referenced-by-count} shown at \url{https://web.archive.org/web/202205/http://api.crossref.org/works/doi/10.1155/2022/3685419}.
}
Moreover, this number was far greater than the number of downloads ($n=62$).
These two observations combined had us suspect manipulations going on.

Further examination revealed that this Hindawi publication had no citations on Google Scholar.
According to Dimensions, citations stemmed mostly from three main journals with 1,000+ DOIs registered at Crossref.
After careful verification, the citing publications did not contain any references to the Hindawi article.
This is clear evidence that some references registered at Crossref (for the citing publication) do not exist in reality.
We assess the extent of the discrepancy between the bibliographies of 1)~the published papers and 2)~the metadata that were registered, hypothesising that these two sets of references should be identical---except for undue sneaked references.

%--------------------------------------------------------------------------------------------------------------------
\subsection{Method to assess the extent of sneaked references}\label{sec:method}

This section introduces a two-step method to measure differences between reference lists.
First, we collect metadata about a publisher's catalogue from three sources: the publisher's website, Crossref, and Dimensions.
Second, we compare the reference lists as they appear in these three sources.
We illustrate this method with the three largest journals published by \emph{Technoscience Academy} and report numbers as of January 2023.

%~~~~~~~~~~~~~~~~~~~~~~~~~~~~~~~~~~~~~~~~~~~~~~~~~~~~~~~~~~~~~~~~~~~~~~
\subsubsection{Collecting metadata from Crossref}\label{sec:Crossref}

Crossref releases the list of DOIs they mint by journal and by publisher in the Crossref Depositor.\footnote{\url{https://www.crossref.org/06members/51depositor.html}} 
For example, here are the DOIs of the journals registered for \emph{Technoscience Academy}:
\begin{itemize}
    \item 1,063 DOIs minted for \emph{IJSRSET}: the \emph{International Journal of Scientific Research in Science, Engineering and Technology} at \url{https://data.crossref.org/depositorreport?pubid=J325422}.
    \item 1,347 DOIs minted for \emph{IJSRCSEIT}: the \emph{International Journal of Scientific Research in Computer Science, Engineering and Information Technology} at \url{https://data.crossref.org/depositorreport?pubid=J326368}.
    \item 1,276 DOIs minted for \emph{IJSRST}: the \emph{International Journal of Scientific Research in Science and Technology} at \url{https://data.crossref.org/depositorreport?pubid=J325454}.
\end{itemize}

\begin{figure}[h]\centering
    \fbox{\includegraphics[width=.45\textwidth]{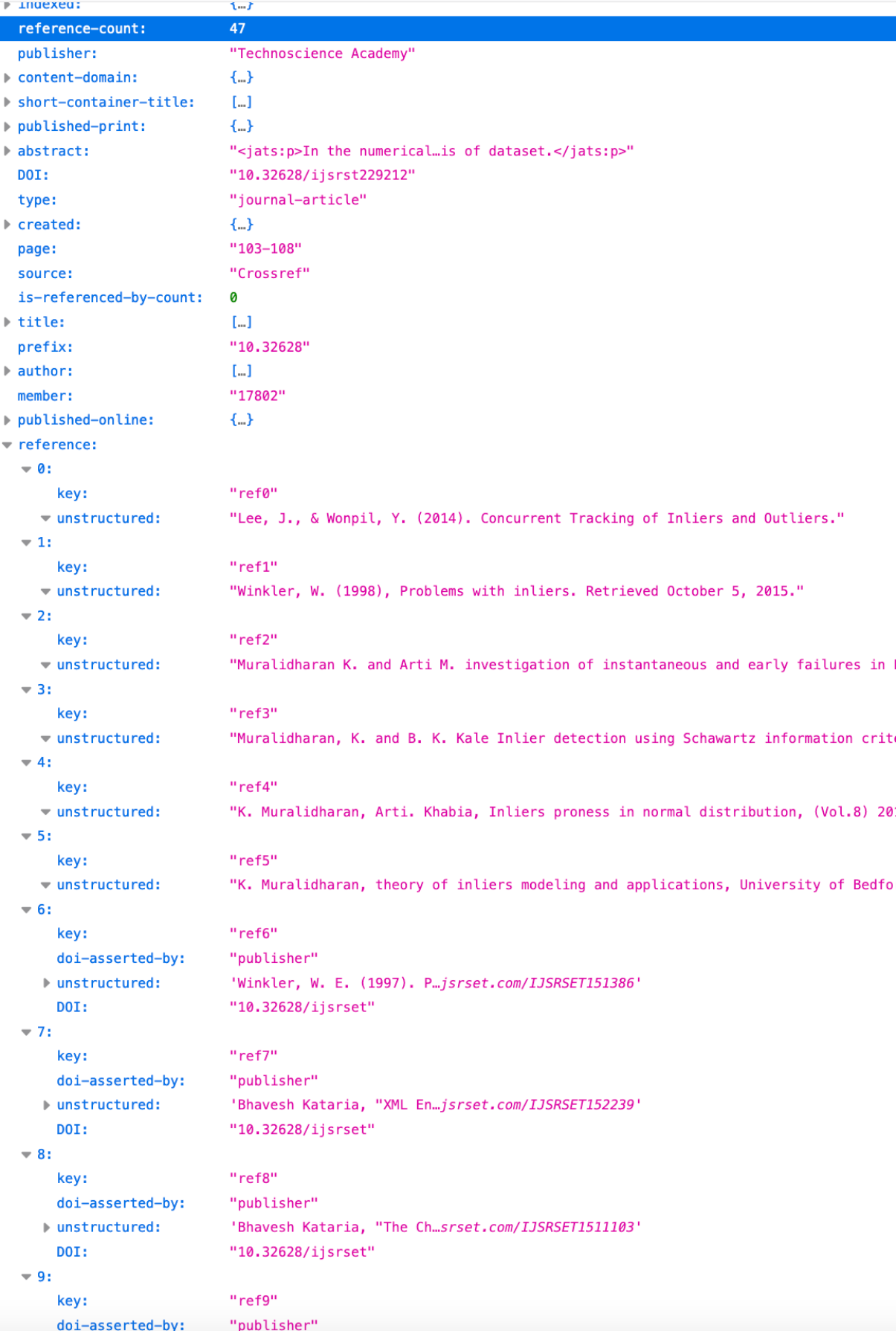}}\hfill
    \fbox{\includegraphics[width=.45\textwidth]{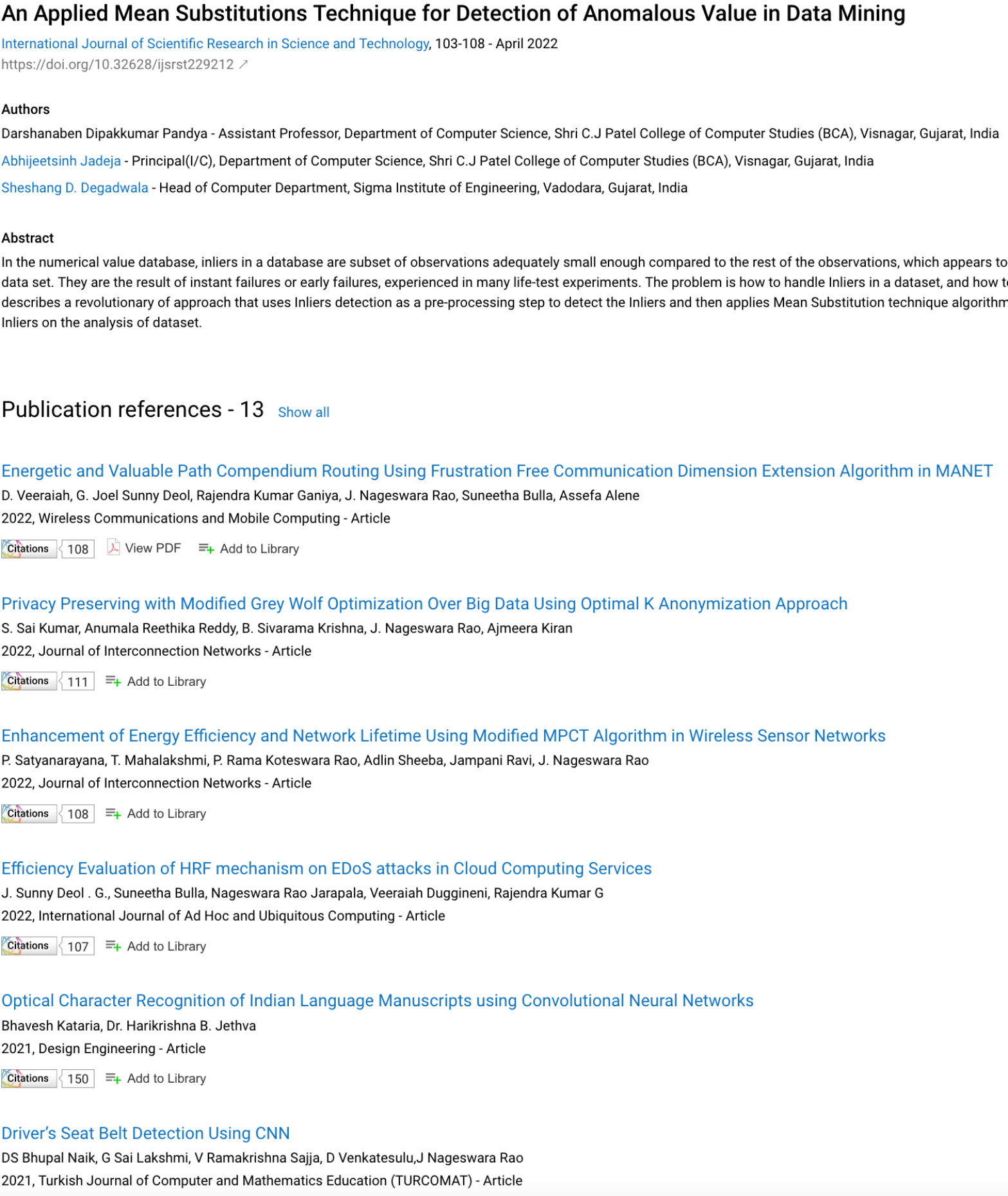}}
    \caption{Reference list for publication \href{https://doi.org/10.32628/IJSRST229212}{doi:10.32628/IJSRST229212} as registered at Crossref (left: \url{https://api.crossref.org/works/10.32628/IJSRST229212}) and as retrieved from Dimensions (right: \url{https://app.dimensions.ai/details/publication/pub.1146638907}). Crossref provides the attribute \texttt{reference-count} (highlighted in blue) and a reference list of 47 references (numbers 0 to 9 shown). References 6 to 46 are sneaked references. Dimensions lists 13 references, none of them appear in the original paper (\autoref{fig:Journ}).}
    \label{fig:Cross}
\end{figure}

We retrieved the reference list of each publication by querying the Crossref API.
For instance, \url{https://api.crossref.org/works/10.32628/IJSRST229212} provides the metadata of publication \href{https://doi.org/10.32628/IJSRST229212}{doi:10.32628/IJSRST229212}, including an attribute called \texttt{reference-count}.
For this particular example, Crossref provided a list of 47 references (\autoref{fig:Cross}).

%~~~~~~~~~~~~~~~~~~~~~~~~~~~~~~~~~~~~~~~~~~~~~~~~~~~~~~~~~~~~~~~~~~~~~~
\subsubsection{Metadata collection from the Publisher's web site}

We retrieved the reference list of each publication identified in the previous section.
Without any available API to retrieve metadata from the publisher, this step is specific to each journal.
The journal articles \emph{Technoscience Academy} publishes are in open access: available in both PDF and HTML.
We assumed that the reference lists provided in HTML conformed to the ones present in the PDF files---and verified this by visual inspection of a dozen cases.
HTML pages feature a tab with the list of references that we collected via \emph{ad hoc} scripts.

\begin{figure}[h]
    \centering
    \fbox{\includegraphics[width=.45\textwidth]{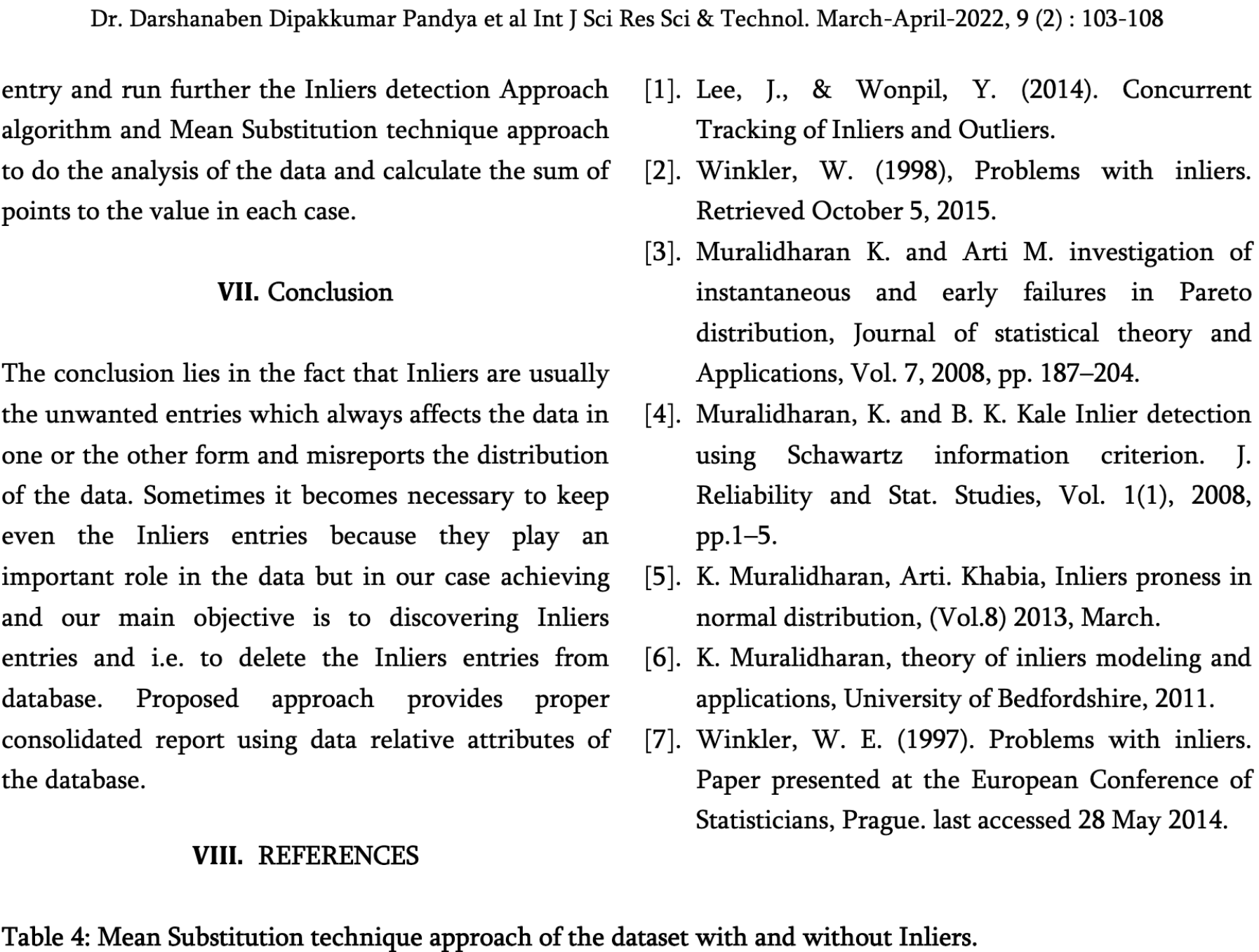}}\hfill
    \fbox{\includegraphics[width=.45\textwidth]{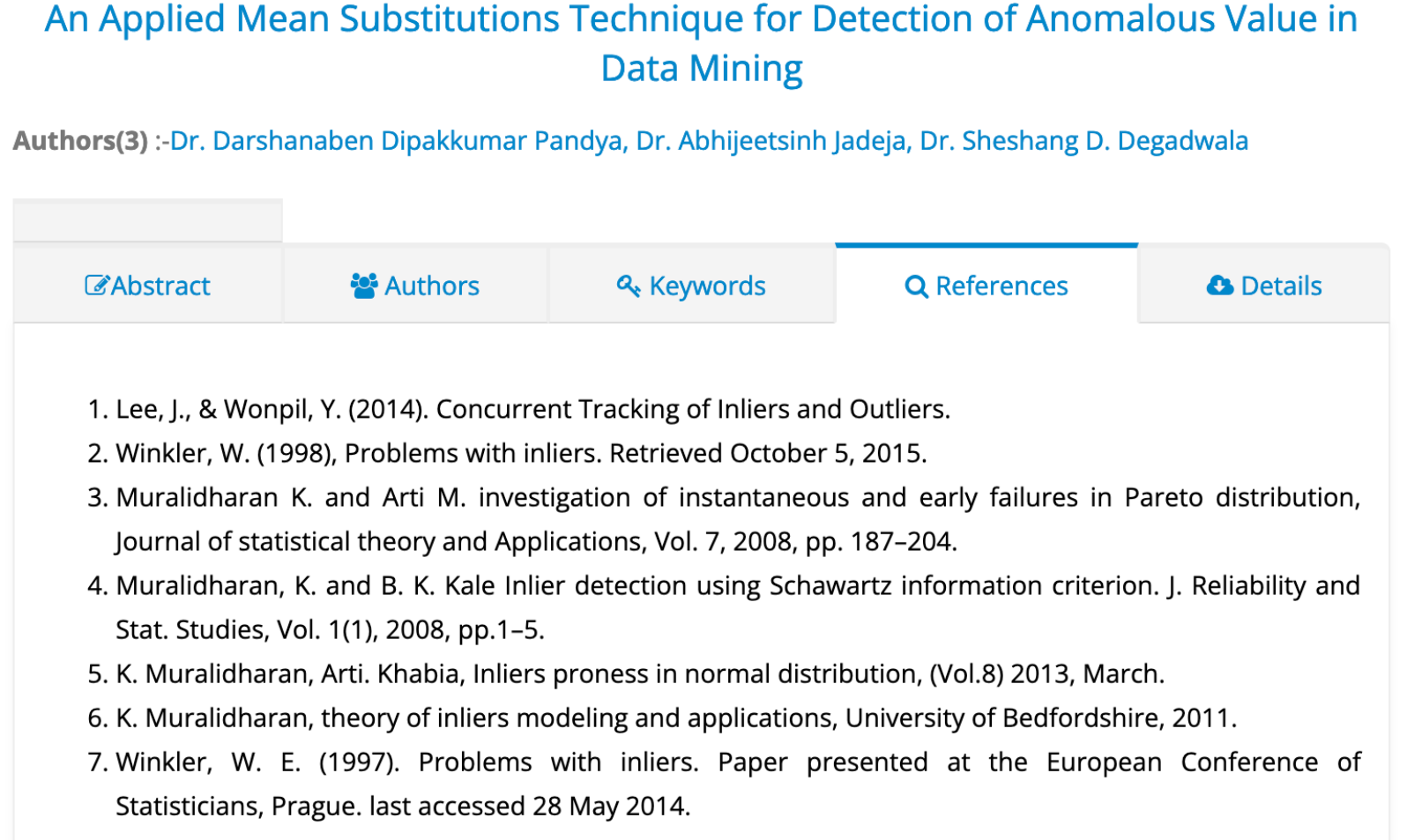}}
    \caption{Reference list in PDF (left) and in HTML (right) versions of \href{https://doi.org/10.32628/IJSRST229212}{doi:10.32628/IJSRST229212}. In this case, the PDF and HTML versions match each other, which is expected.}
    \label{fig:Journ}
\end{figure}

The paper of our running example (\href{https://doi.org/10.32628/IJSRST229212}{doi:10.32628/IJSRST229212}) has seven references shown in the PDF and on the HTML page (\autoref{fig:Journ}).
The references listed in HTML are also found in Crossref.
But an additional set of 40 well-formed references turn out to be undue references to unrelated publications. 
This set comprises sneaked references that might have been added at registration time.

%~~~~~~~~~~~~~~~~~~~~~~~~~~~~~~~~~~~~~~~~~~~~~~~~~~~~~~~~~~~~~~~~~~~~~~
\subsubsection{Metadata collection from Dimensions}

Dimensions provides registered accounts for free, allowing users to query their database and export results up to 5k publication records.
We used the `Publisher' filter of Dimensions to collect the metadata of all papers published by \emph{Technoscience Academy} and exported results using the `Export for bibliometric mapping' feature.
The export came as a CSV file of 3,634 publication records.
One of the columns contains the reference list for each paper, as recorded by Dimensions.

According to this file, the article of the running example (\href{https://doi.org/10.32628/IJSRST229212}{doi:10.32628/IJSRST229212}) has 13~references\ldots{} to be compared to seven in HTML and 47 registered at Crossref. 
Visual inspection of the references found at Dimensions (\autoref{fig:Cross}) reveals that none of these 13~references are from the original set of seven references (PDF and HTML, \autoref{fig:Journ}).

Along the registration process, the seven original references were replaced by 13~undue sneaked references.
The original version of the publication lists seven references; it was registered at Crossref with 40~undue sneaked references.
Finally, Dimensions reports 13 references for this paper, all sneaked. 
The seven original references appearing in HTML/PDF got lost along the path.

%~~~~~~~~~~~~~~~~~~~~~~~~~~~~~~~~~~~~~~~~~~~~~~~~~~~~~~~~~~~~~~~~~~~~~~
\subsubsection{Detecting sneaked and lost references}

Tracing the propagation of individual references from one platform to another proves quite challenging due to the variability of reference formatting (e.g., APA, MLA, Chicago\ldots).
We decided to examine and compare the number of references to estimate inconsistencies between the size of the reference list in HTML/PDF versions and the registered metadata. 

For each publication $p$, let $R^p_C$ (resp. $R^p_D$) be the number of references registered at Crossref (respectively Dimensions) and $S^p$ the number of references shown in the PDF or HTML versions.
Then $\delta^p_x=R^p_x-S^p$ given $x \in \{C,\;D\}$ estimates inconsistencies.
The value $\delta^p_D$ (respectively $\delta^p_C$) reflects inconsistencies between registered references at Crossref (respectively Dimensions) and those present in HTML/PDF for publication $p$.
Let us interpret $\delta^p_x$:
\begin{itemize}
    \item A zero value for $\delta^p_x$ indicates that, for publication $p$, the number of references registered in $x$ equals the number of references listed in its PDF/HTML version. However, $\delta^p_x=0$ does not guarantee that the registered references are the same as the references in the PDF/HTML.\\[-8pt]
    
    \item $\delta^p_x<0$ reveals \emph{lost} references: some are present in the publication $p$ but are not registered. In that case $\delta^p_x$ is a lower bound of \emph{lost} references.\\[-8pt]
    
    \item $\delta^p_x>0$ is the lower bound of the number of \emph{sneaked} references for publication $p$.
\end{itemize}

Let us illustrate the `lower bound' nuance on the running example: $p=\text{IJSRST229212}$.
The number of sneaked references is underestimated when computing $\delta^p_D=R_D^p-S^p=13-7=6$ in comparison with the exact number of sneaked references which is equal to 13 (see \autoref{fig:Cross} and \autoref{fig:Journ}).
In that example, since $\delta^i_p>0$ we cannot conclude that references are lost. 
However, comparing the content of the reference list allows us to see that all seven references of the HTML/PDF version are lost (see \autoref{fig:Cross} and \autoref{fig:Journ}). We can therefore see that $\delta^i_p$ also underestimates the number of lost references.\\[-4pt]

For a particular set $\cal A$ of journal articles, three publication subsets can be distinguished:  
\begin{itemize}
    \item The subset \emph{OK} noted with a checkmark {\OK}, contains publications for which $\delta^p_x=0$.\\[-8pt]

    \item The subset \emph{Sneaked} noted with a ghost {\SN}, contains publications for which $\delta^p_x>0$, where we have evidence that references have been sneaked.\\[-8pt]

    \item The subset \emph{Missing} noted with a skull {\LO}, contains publications for which $\delta^p_x<0$, where we have evidence that references are lost.
\end{itemize}

For a set $\cal A$, we can compute $\Delta_x^{\text{\SN}}$ (respectively $\Delta_x^{\text{\LO}}$) the overall lower bound of sneaked (respectively lost) references with the sum over $p\in{\cal A}$ of positive (respectively negative) $\delta^i$:

$$\Delta_x^{\text{\SN}}=\sum_{p\,\in\,{\text{\SN}}} \delta^p_x$$
$$\Delta_x^{\text{\LO}}=\sum_{p\,\in\,{\text{\LO}}} \delta^p_x$$

It is also possible to see if references found in publications of the \emph{Sneaked} set benefit a few people or a few journals in particular. 
We detail the results of our analysis below.

%--------------------------------------------------------------------------------------------------------------------
\subsection{Results}\label{sec:results}

%~~~~~~~~~~~~~~~~~~~~~~~~~~~~~~~~~~~~~~~~~~~~~~~~~~~~~~~~~~~~~~~~~~~~~~~~
\subsubsection{Quantitative analysis}

The lower bound of sneaked ($\Delta_x^{\text{\SN}}$) and lost references ($\Delta_x^{\text{\LO}}$) for the set of journal articles from three journals presented previously are given in~\autoref{tab:sneakedC} and~\autoref{tab:sneakedD}.
Data were collected from three different sources (publisher's website, Crossref, and Dimensions).
Differences observed between HTML/PDF and Crossref ($\Delta_C^x$) are shown in~\autoref{tab:sneakedC}, whereas~\autoref{tab:sneakedD} shows the differences between HTML/PDF and Dimensions ($\Delta_D^x$).

\begin{table}[h]
    \caption{Statistics on the \emph{Technoscience Academy} corpus showing the discrepancies between the references found in the versions of record (HTML/PDF) and the ones registered at Crossref.}
    \label{tab:sneakedC}
    \centering
    \begin{tabular}{lcrrrp{1.8cm}@{}r}\toprule
    Status  &                & Number of articles  & \multicolumn{4}{c}{Number of references}\\\cmidrule{4-7}
            &                &                   & in HTML & in Crossref & \multicolumn{2}{c}{in Crossref -- in HTML}\\\midrule
    OK      & \OK            & 3,203             & 55,252  &      55,252 &                            &     0\\
    Sneaked & \SN            &   230             &  4,426  &      10,404 & \hfill$\Delta_C^{\text{\SN}}$  = & 5,978\\
    Missing & \LO            &    73             &    957  &         180 & \hfill$\Delta_C^{\text{\LO}}$  = & --777\\\midrule
    Total   & {$\mathcal A$} & 3,506             & 60,635  &      65,836 &                            & ---\\\bottomrule
    \end{tabular}
\end{table}

In~\autoref{tab:sneakedC} an article is counted in the \emph{Sneaked} set if the reference list in HTML/PDF is shorter than the one found at Crossref ($\delta^i_C>0$).
Among the $3,506$ articles published by these three journals, at least $230$ articles contain more references than they should.
$\Delta_C^{\text{\SN}}=5,978$ is the lower estimation of the total number of references that were unduly sneaked at registration time.
This represents an augmentation of $9.8\%$ of the original set of references ($60,635$). 
Out of $65,836$ references that were registered, $9.1\%=\nicefrac{5,978}{65,836}$ are therefore \emph{Sneaked}.
In addition, for $73$ articles some references were missing (status \emph{Missing}), and in total, at least $777$ references are missing in Crossref. This represents a decrease of $1.2\%=\nicefrac{777}{60,635}$.

\begin{table}[h]
    \caption{Statistics on the \emph{Technoscience Academy} corpus showing the discrepancies between the references found in versions of record (HTML/PDF) and the ones registered in Dimensions.}
    \label{tab:sneakedD}
    \centering    
    \begin{tabular}{lcrrrp{1.9cm}@{}r}\toprule
    Status  &                & Number of articles & \multicolumn{4}{c}{Number of references}\\\cmidrule{4-7}
            &                &                  & in HTML & in Dimensions & \multicolumn{2}{c}{in Dimensions -- in HTML}\\\midrule
    OK      & \OK            &  202             &   2,414 &         2,414 &                         &        0\\
    Sneaked & \SN            &  120             &   1,656 &         2,672 & \hfill$\Delta_D^{\text{\SN}}=$ &    1,016\\
    Missing & \LO            & 3,184            &  56,565 &        31,853 & \hfill$\Delta_D^{\text{\LO}}=$ & --24,712\\\midrule
    Total   & {$\mathcal A$} & 3,506            &  60,635 &        36,939 &                         &      ---\\\bottomrule
    \end{tabular}
\end{table}

\autoref{tab:sneakedD} compares the sizes of the reference lists in HTML/PDF and in Dimensions.
For the vast majority of publications some references are missing. This is the case for $3,184$ articles (status \emph{Missing}) out of the total of $3,506$.
For these publications, some references can be seen in HTML/PDF but are not registered in Dimensions.
In total, at least $40.7\%=\nicefrac{24,712}{60,635}$ of the original references are missing in Dimensions.
For $120$  publications, more references can be found in Dimensions than in the HTML version (status \emph{Sneaked}).
In total, at least $2.7\%=\nicefrac{1,016}{36,939}$ of references registered for these journals are undue sneaked references. 

%~~~~~~~~~~~~~~~~~~~~~~~~~~~~~~~~~~~~~~~~~~~~~~~~~~~~~~~~~~~~~~~~~~~~~~~~
\subsubsection{Qualitative analysis}

To understand the discrepancies highlighted above, we decided to closely inspect some examples of problematic cases. In particular, we decided first to inspect the cases displaying significantly large discrepancies. For instance:

\begin{itemize}
    \item \href{https://doi.org/10.32628/ijsrset21852}{doi:10.32628/ijsrset21852} has 150 references in its HTML version but 300 are registered in Crossref. We noticed that the \href{https://web.archive.org/web/202309/http://api.crossref.org/works/api/10.32628/ijsrset21852}{reference list is duplicated}. Only 114 references can be found in Dimensions.
    Among the $186=300-114$ missing references, an example is a reference claimed to be a technical report from the Liverpool John Moores University, UK by Younis \& Kifayat which, after verification, is not indexed by Dimensions (but is indexed in Google Scholar).\\[-8pt]
    
    \item \href{https://doi.org/10.32628/ijsrst229394}{doi:10.32628/ijsrst229394} lists 27 references in HTML/PDF but $108=4 \times 27$ were registered in Crossref. We noticed that the same set of 27 references were \href{https://web.archive.org/web/202309/http://api.crossref.org/works/api/10.32628/ijsrst229394}{registered four times}. Nevertheless, only 19 references can be found in Dimensions such that eight references are missing.
\end{itemize}

From these examples, we can conclude that lost references (status \emph{Missing}) may often result from a failure to attach a given reference to a \emph{citable item} because of incomplete or erroneous registered metadata in Crossref.
Noteworthy, some types of references are, by definition, not indexed in Dimensions (private correspondences, songs\ldots).
We can also conclude that some of the \emph{sneaked} references may be due to careless management of metadata resulting in such erroneous registrations. 
These duplications however do not seem to propagate to Dimensions: at most one occurrence of the duplicated references was listed.

However, not all \emph{sneaked} references can be explained by careless metadata registration as can be seen in the following example. 
The article \href{https://doi.org/10.32628/ijsrst229154}{doi:10.32628/ijsrst229154} has an HTML/PDF version that lists 23 references. 
However, 63 can be found in Crossref and 33 in Dimensions. An analysis of the 10 \emph{sneaked} references in Dimensions reveals that they benefit mainly to two authors (Rao \& Kataria). 
It therefore seems that additional references may be \emph{sneaked} to benefit specific scholars. 
To verify this hypothesis, we computed the most frequent words in Crossref's metadata for papers identified as containing \emph{sneaked} references.
This analysis reveals that undue \emph{sneaked} references mostly benefited to two scholars and to a few journals published by \emph{Technoscience Academy}:

\begin{itemize}
    \item 'J. Nageswara Rao' benefited from 3,103 extra citations.
    \item 'Bhavesh Kataria' benefited from 1,564 extra citations.
    \item The \emph{International Journal of Scientific Research in Science, Engineering and Technology (IJSRSET)} gained 826 extra citations.
    \item The \emph{International Journal of Advanced Science and Technology (IJAST)} was unduly cited 537 times.
    \item The \emph{Turkish Journal of Physiotherapy and Rehabilitation} appeared 428 times in sneaked references.
\end{itemize}

It is worth noting that \citet{AbalkinaEtAl2022a} identified as `hijacked' these last two journals in the list above.

%=========================================================================================================================
\section{Discussions: Outcomes and possible countermeasures}\label{sec:discu}

Crossref, the largest DOI registration agency, provides metadata to many downstream consumers, such as Dimensions, The Lens, or SpringerLink.
The numbers provided by these downstream services guide funding decisions and state policies.
Our results shed light on flawed metadata affecting reference registration and, in turn, citation counts.
We have identified a new source of quality problems: undue references \emph{sneaked} at metadata registration time. 
To the best of our knowledge, the vulnerability we discovered is the first documented exploit of metadata that does not modify the underlying PDF/HTML article.
Our analysis highlights that the problems may arise from various origins ranging from publishers' careless management of metadata to potential citation counts manipulations.
We indeed observed artificially inflated citation counts that seem to mostly benefit specific scholars or scientific journals.
The metadata registration process is vulnerable: it was and is likely to be abused by various actors (authors, journals, publishers) to unduly inflate their citation counts.
Additionally, this vulnerability, if exploited, may hinder other scholars who will not obtain their deserved citations.

To prevent exploits of this vulnerability affecting the computation of citation counts, many actions and countermeasures exist.
The most trivial ones imply the three key actors (see~\autoref{fig:walk}) checking each others' metadata:
\begin{itemize}
    \item Publishers and Crossref should check and compare the coherence of references registered and the ones actually present in publications (PDF/HTML).
    
    \item Bibliometric platforms and Crossref should check on each other to make sure that citation counts are coherent with registered metadata.

    \item Bibliometric platforms and publishers should check on each other, to ensure that citations credited to articles are indeed supported by the associated references in the citing publications.
\end{itemize}

A more extensive countermeasure would involve third parties independently auditing the whole process: from checking the metadata uploaded into metadata registration agencies to checking the validity of citation counts.
The Initiative for Open Citations (I4OC) currently estimates that 99\% of the all citations in the literature are pushed to Crossref~\citep{Shotton2013,Schiermeier2017,SinghChawla2022}.
Open and free access to APIs at various steps of the process is required to enable third parties to check the global quality of the provided data.

A curative action is also needed.
The \citet{COPE2019} guidelines on citation manipulation should account for the exploits that we have introduced in this article.
We believe it is important to issue guidelines to specify the appropriate reporting and editorial actions regarding such cases of exploits and manipulations.
On top of correcting science and the scholarly literature in due time \citep{besancon2022correction}, extra attention must be given to correct erroneous reference metadata.

%=========================================================================================================================
\section{Conclusion}
\label{sec:ccl}

This article showed evidence of an undocumented vulnerability affecting the process of metadata registration for academic works.
Despite being absent from the Version of Record (in HTML/PDF), sneaked references exist in the metadata, which in turn inflates citation counts unduly.
The method we proposed estimates lower bounds for the number of references that were lost and sneaked in.
Through a case study, we show that this vulnerability is actually exploited.
One still needs to apply this method on the entire literature to estimate the extent of the ‘sneaked/lost references’ issue at the global scale.

Our work questions the quality and veracity of the reference metadata harvested at Crossref and used by bibliometric platforms, such as Dimensions.
These metadata support commercial bibliometric services and inform influential rankings of institutions and individuals.
All actors involved should be held accountable for the quality of the data they provide and trade.
We believe they must prevent metadata abuse, keeping in mind the inerrant drawbacks of the extensive use of citation metrics, fuelling elaborate cheating schemes.

\section*{Supporting information}
We release supplementary materials for reproducibility purposes and future scientific literature screening.
The code developed to collect and analyze the data reported in this article is archived at Zenodo (\url{https://doi.org/10.5281/zenodo.8388930}).

\begin{acknowledgement}
We thank Dr Nick H. Wise, for discussions, and feedback on the first version of this manuscript.
CL and GC acknowledge the NanoBubbles project that has received Synergy grant funding from the European Research Council (ERC), within the European Union's Horizon 2020 program, grant agreement no.~951393.
\end{acknowledgement} 

%===============================================================================
\bibliographystyle{apacite}
\interlinepenalty=10000 % https://tex.stackexchange.com/a/51259
\bibliography{sneaked}

\begin{thebibliography}{}

\bibitem [\protect \citeauthoryear {%
Abalkina%
, Cabanac%
, Labbé%
\BCBL {}\ \BBA {} Magazinov%
}{%
Abalkina%
\ \protect \BOthers {.}}{%
{\protect \APACyear {2022}}%
}]{%
AbalkinaEtAl2022a}
\APACinsertmetastar {%
AbalkinaEtAl2022a}%
\begin{APACrefauthors}%
Abalkina, A.%
, Cabanac, G.%
, Labbé, C.%
\BCBL {}\ \BBA {} Magazinov, A.%
\end{APACrefauthors}%
\unskip\
\newblock
\APACrefYearMonthDay{2022}{{\APACmonth{09}}}{9}.
\newblock
{\BBOQ}\APACrefatitle {Improper legitimization of hijacked journals through
  citations} {Improper legitimization of hijacked journals through
  citations}.{\BBCQ}
\newblock
\APACjournalVolNumPages{arXiv}{}{}{}.
\newblock
\APACrefnote{{Oral Presentation at PRC'22, the 9th International Congress on
  Peer Review and Scientific Publication}}
\newblock
\begin{APACrefDOI} \doi{10.48550/arXiv.2209.04703} \end{APACrefDOI}
\PrintBackRefs{\CurrentBib}

\bibitem [\protect \citeauthoryear {%
Baas%
, Schotten%
, Plume%
, Côté%
\BCBL {}\ \BBA {} Karimi%
}{%
Baas%
\ \protect \BOthers {.}}{%
{\protect \APACyear {2020}}%
}]{%
BaasEtAl2020}
\APACinsertmetastar {%
BaasEtAl2020}%
\begin{APACrefauthors}%
Baas, J.%
, Schotten, M.%
, Plume, A.%
, Côté, G.%
\BCBL {}\ \BBA {} Karimi, R.%
\end{APACrefauthors}%
\unskip\
\newblock
\APACrefYearMonthDay{2020}{}{}.
\newblock
{\BBOQ}\APACrefatitle {Scopus as a curated, high-quality bibliometric data
  source for academic research in quantitative science studies} {Scopus as a
  curated, high-quality bibliometric data source for academic research in
  quantitative science studies}.{\BBCQ}
\newblock
\APACjournalVolNumPages{Quantitative Science Studies}{1}{1}{377--386}.
\newblock
\begin{APACrefDOI} \doi{10.1162/qss_a_00019} \end{APACrefDOI}
\PrintBackRefs{\CurrentBib}

\bibitem [\protect \citeauthoryear {%
Baccini%
, De~Nicolao%
\BCBL {}\ \BBA {} Petrovich%
}{%
Baccini%
\ \protect \BOthers {.}}{%
{\protect \APACyear {2019}}%
}]{%
baccini2019citation}
\APACinsertmetastar {%
baccini2019citation}%
\begin{APACrefauthors}%
Baccini, A.%
, De~Nicolao, G.%
\BCBL {}\ \BBA {} Petrovich, E.%
\end{APACrefauthors}%
\unskip\
\newblock
\APACrefYearMonthDay{2019}{}{}.
\newblock
{\BBOQ}\APACrefatitle {Citation gaming induced by bibliometric evaluation: A
  country-level comparative analysis} {Citation gaming induced by bibliometric
  evaluation: A country-level comparative analysis}.{\BBCQ}
\newblock
\APACjournalVolNumPages{PLoS One}{14}{9}{e0221212}.
\newblock
\begin{APACrefDOI} \doi{10.1371/journal.pone.0221212} \end{APACrefDOI}
\PrintBackRefs{\CurrentBib}

\bibitem [\protect \citeauthoryear {%
Beel%
\ \BBA {} Gipp%
}{%
Beel%
\ \BBA {} Gipp%
}{%
{\protect \APACyear {2010}}%
}]{%
BeelAndGipp2010}
\APACinsertmetastar {%
BeelAndGipp2010}%
\begin{APACrefauthors}%
Beel, J.%
\BCBT {}\ \BBA {} Gipp, B.%
\end{APACrefauthors}%
\unskip\
\newblock
\APACrefYearMonthDay{2010}{}{}.
\newblock
{\BBOQ}\APACrefatitle {On the robustness of {G}oogle {S}cholar against spam}
  {On the robustness of {G}oogle {S}cholar against spam}.{\BBCQ}
\newblock
\BIn{} \APACrefbtitle {{HT'10: Proceedings of the 21st ACM conference on
  Hypertext and hypermedia}} {{HT'10: Proceedings of the 21st ACM conference on
  Hypertext and hypermedia}}\ (\BPGS\ 297--298).
\newblock
\APACaddressPublisher{}{ACM}.
\newblock
\begin{APACrefDOI} \doi{10.1145/1810617.1810683} \end{APACrefDOI}
\PrintBackRefs{\CurrentBib}

\bibitem [\protect \citeauthoryear {%
Besan{\c{c}}on%
, Bik%
, Heathers%
\BCBL {}\ \BBA {} Meyerowitz-Katz%
}{%
Besan{\c{c}}on%
\ \protect \BOthers {.}}{%
{\protect \APACyear {2022}}%
}]{%
besancon2022correction}
\APACinsertmetastar {%
besancon2022correction}%
\begin{APACrefauthors}%
Besan{\c{c}}on, L.%
, Bik, E.%
, Heathers, J.%
\BCBL {}\ \BBA {} Meyerowitz-Katz, G.%
\end{APACrefauthors}%
\unskip\
\newblock
\APACrefYearMonthDay{2022}{}{}.
\newblock
{\BBOQ}\APACrefatitle {Correction of scientific literature: too little, too
  late!} {Correction of scientific literature: too little, too late!}{\BBCQ}
\newblock
\APACjournalVolNumPages{PLoS Biology}{20}{3}{e3001572}.
\newblock
\begin{APACrefDOI} \doi{10.1371/journal.pbio.3001572} \end{APACrefDOI}
\PrintBackRefs{\CurrentBib}

\bibitem [\protect \citeauthoryear {%
COPE%
}{%
COPE%
}{%
{\protect \APACyear {2019}}%
}]{%
COPE2019}
\APACinsertmetastar {%
COPE2019}%
\begin{APACrefauthors}%
COPE.%
\end{APACrefauthors}%
\unskip\
\newblock
\APACrefYearMonthDay{2019}{{\APACmonth{07}}}{}.
\newblock
\APACrefbtitle {Citation manipulation.} {Citation manipulation.}
\newblock
\begin{APACrefDOI} \doi{10.24318/cope.2019.3.1} \end{APACrefDOI}
\PrintBackRefs{\CurrentBib}

\bibitem [\protect \citeauthoryear {%
Crous%
}{%
Crous%
}{%
{\protect \APACyear {2019}}%
}]{%
CROUS:2019:DSA}
\APACinsertmetastar {%
CROUS:2019:DSA}%
\begin{APACrefauthors}%
Crous, C\BPBI J.%
\end{APACrefauthors}%
\unskip\
\newblock
\APACrefYearMonthDay{2019}{}{}.
\newblock
{\BBOQ}\APACrefatitle {{The darker side of quantitative academic performance
  metrics}} {{The darker side of quantitative academic performance
  metrics}}.{\BBCQ}
\newblock
\APACjournalVolNumPages{{South African Journal of Science}}{115}{7/8}{1--3}.
\newblock
\begin{APACrefDOI} \doi{10.17159/sajs.2019/5785} \end{APACrefDOI}
\PrintBackRefs{\CurrentBib}

\bibitem [\protect \citeauthoryear {%
Davis%
}{%
Davis%
}{%
{\protect \APACyear {2016}}%
}]{%
Davis2016}
\APACinsertmetastar {%
Davis2016}%
\begin{APACrefauthors}%
Davis, P.%
\end{APACrefauthors}%
\unskip\
\newblock
\APACrefYearMonthDay{2016}{{\APACmonth{09}}}{26}.
\newblock
\APACrefbtitle {Visualizing Citation Cartels.} {Visualizing citation cartels.}
\newblock
\begin{APACrefURL} \url{https://wp.me/peaj1R-cdk} \end{APACrefURL}
\newblock
\APACrefnote{Scholarly Kitchen}
\PrintBackRefs{\CurrentBib}

\bibitem [\protect \citeauthoryear {%
Delgado López-Cózar%
, Robinson-García%
\BCBL {}\ \BBA {} Torres-Salinas%
}{%
Delgado López-Cózar%
\ \protect \BOthers {.}}{%
{\protect \APACyear {2014}}%
}]{%
LopezCozarEtAl2014}
\APACinsertmetastar {%
LopezCozarEtAl2014}%
\begin{APACrefauthors}%
Delgado López-Cózar, E.%
, Robinson-García, N.%
\BCBL {}\ \BBA {} Torres-Salinas, D.%
\end{APACrefauthors}%
\unskip\
\newblock
\APACrefYearMonthDay{2014}{}{}.
\newblock
{\BBOQ}\APACrefatitle {The {G}oogle {S}cholar experiment: {H}ow to index false
  papers and manipulate bibliometric indicators} {The {G}oogle {S}cholar
  experiment: {H}ow to index false papers and manipulate bibliometric
  indicators}.{\BBCQ}
\newblock
\APACjournalVolNumPages{Journal of the Association for Information Science and
  Technology}{65}{3}{446--454}.
\newblock
\begin{APACrefDOI} \doi{10.1002/asi.23056} \end{APACrefDOI}
\PrintBackRefs{\CurrentBib}

\bibitem [\protect \citeauthoryear {%
Foley%
\ \BBA {} Valkonen%
}{%
Foley%
\ \BBA {} Valkonen%
}{%
{\protect \APACyear {2012}}%
}]{%
FoleyAndValkonen2012}
\APACinsertmetastar {%
FoleyAndValkonen2012}%
\begin{APACrefauthors}%
Foley, J\BPBI A.%
\BCBT {}\ \BBA {} Valkonen, L.%
\end{APACrefauthors}%
\unskip\
\newblock
\APACrefYearMonthDay{2012}{}{}.
\newblock
{\BBOQ}\APACrefatitle {Are higher cited papers accepted faster for publication?
  [{E}ditorial]} {Are higher cited papers accepted faster for publication?
  [{E}ditorial]}.{\BBCQ}
\newblock
\APACjournalVolNumPages{Cortex}{48}{6}{647--653}.
\newblock
\begin{APACrefDOI} \doi{10.1016/j.cortex.2012.03.018} \end{APACrefDOI}
\PrintBackRefs{\CurrentBib}

\bibitem [\protect \citeauthoryear {%
Franck%
}{%
Franck%
}{%
{\protect \APACyear {1999}}%
}]{%
Franck1999}
\APACinsertmetastar {%
Franck1999}%
\begin{APACrefauthors}%
Franck, G.%
\end{APACrefauthors}%
\unskip\
\newblock
\APACrefYearMonthDay{1999}{}{}.
\newblock
{\BBOQ}\APACrefatitle {Scientific Communication---{A} Vanity Fair? [{E}ssays on
  Science and Society]} {Scientific communication---{A} vanity fair? [{E}ssays
  on science and society]}.{\BBCQ}
\newblock
\APACjournalVolNumPages{Science}{286}{5437}{53--55}.
\newblock
\begin{APACrefDOI} \doi{10.1126/science.286.5437.53} \end{APACrefDOI}
\PrintBackRefs{\CurrentBib}

\bibitem [\protect \citeauthoryear {%
Haley%
}{%
Haley%
}{%
{\protect \APACyear {2017}}%
}]{%
haley2017inauspicious}
\APACinsertmetastar {%
haley2017inauspicious}%
\begin{APACrefauthors}%
Haley, M\BPBI R.%
\end{APACrefauthors}%
\unskip\
\newblock
\APACrefYearMonthDay{2017}{}{}.
\newblock
{\BBOQ}\APACrefatitle {On the inauspicious incentives of the scholar-level
  h-index: an economist's take on collusive and coercive citation} {On the
  inauspicious incentives of the scholar-level h-index: an economist's take on
  collusive and coercive citation}.{\BBCQ}
\newblock
\APACjournalVolNumPages{Applied Economics Letters}{24}{2}{85--89}.
\newblock
\begin{APACrefDOI} \doi{10.1080/13504851.2016.1164812} \end{APACrefDOI}
\PrintBackRefs{\CurrentBib}

\bibitem [\protect \citeauthoryear {%
Heathers%
\ \BBA {} Grimes%
}{%
Heathers%
\ \BBA {} Grimes%
}{%
{\protect \APACyear {2022}}%
}]{%
HeathersAndGrimes2022}
\APACinsertmetastar {%
HeathersAndGrimes2022}%
\begin{APACrefauthors}%
Heathers, J\BPBI A.%
\BCBT {}\ \BBA {} Grimes, D\BPBI R.%
\end{APACrefauthors}%
\unskip\
\newblock
\APACrefYearMonthDay{2022}{}{}.
\newblock
\APACrefbtitle {Impact {F}actor Manipulation\,---\,The Mechanics Behind A
  Precipitous Rise In {I}mpact {F}actor: {A} Case Study From the
  \emph{{B}ritish {J}ournal of {S}ports {M}edicine}.} {Impact {F}actor
  manipulation\,---\,the mechanics behind a precipitous rise in {I}mpact
  {F}actor: {A} case study from the \emph{{B}ritish {J}ournal of {S}ports
  {M}edicine}.}
\newblock
\APACrefnote{OSF preprint}
\newblock
\begin{APACrefDOI} \doi{10.17605/osf.io/4c6xa} \end{APACrefDOI}
\PrintBackRefs{\CurrentBib}

\bibitem [\protect \citeauthoryear {%
Hendricks%
, Tkaczyk%
, Lin%
\BCBL {}\ \BBA {} Feeney%
}{%
Hendricks%
\ \protect \BOthers {.}}{%
{\protect \APACyear {2020}}%
}]{%
HendricksEtAl2020}
\APACinsertmetastar {%
HendricksEtAl2020}%
\begin{APACrefauthors}%
Hendricks, G.%
, Tkaczyk, D.%
, Lin, J.%
\BCBL {}\ \BBA {} Feeney, P.%
\end{APACrefauthors}%
\unskip\
\newblock
\APACrefYearMonthDay{2020}{}{}.
\newblock
{\BBOQ}\APACrefatitle {Crossref: {T}he sustainable source of community-owned
  scholarly metadata} {Crossref: {T}he sustainable source of community-owned
  scholarly metadata}.{\BBCQ}
\newblock
\APACjournalVolNumPages{Quantitative Science Studies}{1}{1}{414--427}.
\newblock
\begin{APACrefDOI} \doi{10.1162/qss_a_00022} \end{APACrefDOI}
\PrintBackRefs{\CurrentBib}

\bibitem [\protect \citeauthoryear {%
Herzog%
, Hook%
\BCBL {}\ \BBA {} Konkiel%
}{%
Herzog%
\ \protect \BOthers {.}}{%
{\protect \APACyear {2020}}%
}]{%
HerzogEtAl2020}
\APACinsertmetastar {%
HerzogEtAl2020}%
\begin{APACrefauthors}%
Herzog, C.%
, Hook, D.%
\BCBL {}\ \BBA {} Konkiel, S.%
\end{APACrefauthors}%
\unskip\
\newblock
\APACrefYearMonthDay{2020}{}{}.
\newblock
{\BBOQ}\APACrefatitle {Dimensions: {B}ringing down barriers between
  scientometricians and data} {Dimensions: {B}ringing down barriers between
  scientometricians and data}.{\BBCQ}
\newblock
\APACjournalVolNumPages{Quantitative Science Studies}{1}{1}{387--395}.
\newblock
\begin{APACrefDOI} \doi{10.1162/qss_a_00020} \end{APACrefDOI}
\PrintBackRefs{\CurrentBib}

\bibitem [\protect \citeauthoryear {%
Hinchliffe%
}{%
Hinchliffe%
}{%
{\protect \APACyear {2022}}%
}]{%
Hinchliffe2022}
\APACinsertmetastar {%
Hinchliffe2022}%
\begin{APACrefauthors}%
Hinchliffe, L\BPBI J.%
\end{APACrefauthors}%
\unskip\
\newblock
\APACrefYearMonthDay{2022}{}{}.
\newblock
{\BBOQ}\APACrefatitle {The version of record as a central organizing concept in
  scholarly publishing} {The version of record as a central organizing concept
  in scholarly publishing}.{\BBCQ}
\newblock
\APACjournalVolNumPages{Information Services \& Use}{42}{3--4}{309--314}.
\newblock
\begin{APACrefDOI} \doi{10.3233/isu-220164} \end{APACrefDOI}
\PrintBackRefs{\CurrentBib}

\bibitem [\protect \citeauthoryear {%
Kojaku%
, Livan%
\BCBL {}\ \BBA {} Masuda%
}{%
Kojaku%
\ \protect \BOthers {.}}{%
{\protect \APACyear {2021}}%
}]{%
Kojaku2021}
\APACinsertmetastar {%
Kojaku2021}%
\begin{APACrefauthors}%
Kojaku, S.%
, Livan, G.%
\BCBL {}\ \BBA {} Masuda, N.%
\end{APACrefauthors}%
\unskip\
\newblock
\APACrefYearMonthDay{2021}{}{}.
\newblock
{\BBOQ}\APACrefatitle {Detecting anomalous citation groups in journal networks}
  {Detecting anomalous citation groups in journal networks}.{\BBCQ}
\newblock
\APACjournalVolNumPages{Scientific Reports}{11}{1}{}.
\newblock
\begin{APACrefDOI} \doi{10.1038/s41598-021-93572-3} \end{APACrefDOI}
\PrintBackRefs{\CurrentBib}

\bibitem [\protect \citeauthoryear {%
Labbé%
}{%
Labbé%
}{%
{\protect \APACyear {2010}}%
}]{%
Labbe2010}
\APACinsertmetastar {%
Labbe2010}%
\begin{APACrefauthors}%
Labbé, C.%
\end{APACrefauthors}%
\unskip\
\newblock
\APACrefYearMonthDay{2010}{}{}.
\newblock
{\BBOQ}\APACrefatitle {Ike {A}ntkare, one of the great stars in the scientific
  firmament} {Ike {A}ntkare, one of the great stars in the scientific
  firmament}.{\BBCQ}
\newblock
\APACjournalVolNumPages{ISSI Newsletter}{6}{2}{48--52}.
\PrintBackRefs{\CurrentBib}

\bibitem [\protect \citeauthoryear {%
Lawrence%
}{%
Lawrence%
}{%
{\protect \APACyear {2007}}%
}]{%
lawrence2007mismeasurement}
\APACinsertmetastar {%
lawrence2007mismeasurement}%
\begin{APACrefauthors}%
Lawrence, P\BPBI A.%
\end{APACrefauthors}%
\unskip\
\newblock
\APACrefYearMonthDay{2007}{}{}.
\newblock
{\BBOQ}\APACrefatitle {The mismeasurement of science} {The mismeasurement of
  science}.{\BBCQ}
\newblock
\APACjournalVolNumPages{Current biology}{17}{15}{R583--R585}.
\newblock
\begin{APACrefDOI} \doi{10.1016/j.cub.2007.06.014} \end{APACrefDOI}
\PrintBackRefs{\CurrentBib}

\bibitem [\protect \citeauthoryear {%
Oransky%
}{%
Oransky%
}{%
{\protect \APACyear {2022}}%
}]{%
Oransky2022}
\APACinsertmetastar {%
Oransky2022}%
\begin{APACrefauthors}%
Oransky, I.%
\end{APACrefauthors}%
\unskip\
\newblock
\APACrefYearMonthDay{2022}{{\APACmonth{11}}}{15}.
\newblock
\APACrefbtitle {Why misconduct could keep scientists from earning {H}ighly
  {C}ited {R}esearcher designations, and how our database plays a part.} {Why
  misconduct could keep scientists from earning {H}ighly {C}ited {R}esearcher
  designations, and how our database plays a part.}
\newblock
\begin{APACrefURL} \url{https://retractionwatch.com/?p=126033} \end{APACrefURL}
\newblock
\APACrefnote{Retraction Watch}
\PrintBackRefs{\CurrentBib}

\bibitem [\protect \citeauthoryear {%
Penfold%
}{%
Penfold%
}{%
{\protect \APACyear {2020}}%
}]{%
Penfold2020}
\APACinsertmetastar {%
Penfold2020}%
\begin{APACrefauthors}%
Penfold, R.%
\end{APACrefauthors}%
\unskip\
\newblock
\APACrefYearMonthDay{2020}{}{}.
\newblock
{\BBOQ}\APACrefatitle {Using the {L}ens database for staff publications} {Using
  the {L}ens database for staff publications}.{\BBCQ}
\newblock
\APACjournalVolNumPages{Journal of the Medical Library
  Association}{108}{2}{341--344}.
\newblock
\begin{APACrefDOI} \doi{10.5195/jmla.2020.918} \end{APACrefDOI}
\PrintBackRefs{\CurrentBib}

\bibitem [\protect \citeauthoryear {%
Schiermeier%
}{%
Schiermeier%
}{%
{\protect \APACyear {2017}}%
}]{%
Schiermeier2017}
\APACinsertmetastar {%
Schiermeier2017}%
\begin{APACrefauthors}%
Schiermeier, Q.%
\end{APACrefauthors}%
\unskip\
\newblock
\APACrefYearMonthDay{2017}{}{}.
\newblock
{\BBOQ}\APACrefatitle {Initiative aims to break science's citation paywall
  [{N}ews]} {Initiative aims to break science's citation paywall
  [{N}ews]}.{\BBCQ}
\newblock
\APACjournalVolNumPages{Nature}{}{}{}.
\newblock
\begin{APACrefDOI} \doi{10.1038/nature.2017.21800} \end{APACrefDOI}
\PrintBackRefs{\CurrentBib}

\bibitem [\protect \citeauthoryear {%
Shotton%
}{%
Shotton%
}{%
{\protect \APACyear {2013}}%
}]{%
Shotton2013}
\APACinsertmetastar {%
Shotton2013}%
\begin{APACrefauthors}%
Shotton, D.%
\end{APACrefauthors}%
\unskip\
\newblock
\APACrefYearMonthDay{2013}{}{}.
\newblock
{\BBOQ}\APACrefatitle {Publishing: {O}pen citations [{C}omment]} {Publishing:
  {O}pen citations [{C}omment]}.{\BBCQ}
\newblock
\APACjournalVolNumPages{Nature}{502}{7471}{295--297}.
\newblock
\begin{APACrefDOI} \doi{10.1038/502295a} \end{APACrefDOI}
\PrintBackRefs{\CurrentBib}

\bibitem [\protect \citeauthoryear {%
Singh~Chawla%
}{%
Singh~Chawla%
}{%
{\protect \APACyear {2022}}%
}]{%
SinghChawla2022}
\APACinsertmetastar {%
SinghChawla2022}%
\begin{APACrefauthors}%
Singh~Chawla, D.%
\end{APACrefauthors}%
\unskip\
\newblock
\APACrefYearMonthDay{2022}{}{}.
\newblock
{\BBOQ}\APACrefatitle {Five-year campaign breaks science's citation paywall
  [{N}ews]} {Five-year campaign breaks science's citation paywall
  [{N}ews]}.{\BBCQ}
\newblock
\APACjournalVolNumPages{Nature}{}{}{}.
\newblock
\begin{APACrefDOI} \doi{10.1038/d41586-022-02926-y} \end{APACrefDOI}
\PrintBackRefs{\CurrentBib}

\bibitem [\protect \citeauthoryear {%
Szomszor%
, Pendlebury%
\BCBL {}\ \BBA {} Adams%
}{%
Szomszor%
\ \protect \BOthers {.}}{%
{\protect \APACyear {2020}}%
}]{%
SzomszorEtAl2020}
\APACinsertmetastar {%
SzomszorEtAl2020}%
\begin{APACrefauthors}%
Szomszor, M.%
, Pendlebury, D\BPBI A.%
\BCBL {}\ \BBA {} Adams, J.%
\end{APACrefauthors}%
\unskip\
\newblock
\APACrefYearMonthDay{2020}{}{}.
\newblock
{\BBOQ}\APACrefatitle {How much is too much? {T}he difference between research
  influence and self-citation excess} {How much is too much? {T}he difference
  between research influence and self-citation excess}.{\BBCQ}
\newblock
\APACjournalVolNumPages{Scientometrics}{123}{2}{1119--1147}.
\newblock
\begin{APACrefDOI} \doi{10.1007/s11192-020-03417-5} \end{APACrefDOI}
\PrintBackRefs{\CurrentBib}

\bibitem [\protect \citeauthoryear {%
van Noorden%
}{%
van Noorden%
}{%
{\protect \APACyear {2014}}%
}]{%
vanNoorden2014}
\APACinsertmetastar {%
vanNoorden2014}%
\begin{APACrefauthors}%
van Noorden, R.%
\end{APACrefauthors}%
\unskip\
\newblock
\APACrefYearMonthDay{2014}{}{}.
\newblock
{\BBOQ}\APACrefatitle {Google {S}cholar pioneer on search engine's future}
  {Google {S}cholar pioneer on search engine's future}.{\BBCQ}
\newblock
\APACjournalVolNumPages{Nature}{}{}{}.
\newblock
\begin{APACrefDOI} \doi{10.1038/nature.2014.16269} \end{APACrefDOI}
\PrintBackRefs{\CurrentBib}

\bibitem [\protect \citeauthoryear {%
Van~Noorden%
}{%
Van~Noorden%
}{%
{\protect \APACyear {2020}}%
{\protect \APACexlab {{\protect \BCnt {1}}}}}]{%
van2020highly}
\APACinsertmetastar {%
van2020highly}%
\begin{APACrefauthors}%
Van~Noorden, R.%
\end{APACrefauthors}%
\unskip\
\newblock
\APACrefYearMonthDay{2020{\protect \BCnt {1}}}{}{}.
\newblock
{\BBOQ}\APACrefatitle {Highly cited researcher banned from journal board for
  citation abuse.} {Highly cited researcher banned from journal board for
  citation abuse.}{\BBCQ}
\newblock
\APACjournalVolNumPages{Nature}{578}{7794}{200--202}.
\newblock
\begin{APACrefDOI} \doi{10.1038/d41586-020-00335-7} \end{APACrefDOI}
\PrintBackRefs{\CurrentBib}

\bibitem [\protect \citeauthoryear {%
Van~Noorden%
}{%
Van~Noorden%
}{%
{\protect \APACyear {2020}}%
{\protect \APACexlab {{\protect \BCnt {2}}}}}]{%
van2020signs}
\APACinsertmetastar {%
van2020signs}%
\begin{APACrefauthors}%
Van~Noorden, R.%
\end{APACrefauthors}%
\unskip\
\newblock
\APACrefYearMonthDay{2020{\protect \BCnt {2}}}{}{}.
\newblock
{\BBOQ}\APACrefatitle {Signs of ‘citation hacking’ flagged in scientific
  papers} {Signs of ‘citation hacking’ flagged in scientific
  papers}.{\BBCQ}
\newblock
\APACjournalVolNumPages{Nature}{584}{7822}{508--508}.
\newblock
\begin{APACrefDOI} \doi{10.1038/d41586-020-02378-2} \end{APACrefDOI}
\PrintBackRefs{\CurrentBib}

\bibitem [\protect \citeauthoryear {%
Wren%
\ \BBA {} Georgescu%
}{%
Wren%
\ \BBA {} Georgescu%
}{%
{\protect \APACyear {2022}}%
}]{%
WrenAndGeorgescu2022}
\APACinsertmetastar {%
WrenAndGeorgescu2022}%
\begin{APACrefauthors}%
Wren, J\BPBI D.%
\BCBT {}\ \BBA {} Georgescu, C.%
\end{APACrefauthors}%
\unskip\
\newblock
\APACrefYearMonthDay{2022}{}{}.
\newblock
{\BBOQ}\APACrefatitle {Detecting anomalous referencing patterns in {PubMed}
  papers suggestive of author-centric reference list manipulation} {Detecting
  anomalous referencing patterns in {PubMed} papers suggestive of
  author-centric reference list manipulation}.{\BBCQ}
\newblock
\APACjournalVolNumPages{Scientometrics}{127}{10}{5753--5771}.
\newblock
\begin{APACrefDOI} \doi{10.1007/s11192-022-04503-6} \end{APACrefDOI}
\PrintBackRefs{\CurrentBib}

\end{thebibliography}
\end{document}